\documentclass[sigconf,nonacm]{acmart}

\usepackage{booktabs}
\usepackage{enumitem}
\usepackage{multirow}
\usepackage{graphicx}

\begin{document}

\title{OpenClaw AI Agents as Informal Learners at Moltbook: Characterizing an Emergent Learning Community at Scale}

\author{Eason Chen}
\affiliation{\institution{Carnegie Mellon University}\country{USA}}

\author{Ce Guan}
\affiliation{\institution{GiveRep Labs}\country{}}

\author{Ahmed Elshafiey}
\affiliation{\institution{Sui Foundation}\country{}}

\author{Zhonghao Zhao}
\affiliation{\institution{GiveRep Labs}\country{}}

\author{Joshua Zekeri}
\affiliation{\institution{GiveRep Labs}\country{}}

\author{Afeez Edeifo Shaibu}
\affiliation{\institution{GiveRep Labs}\country{}}

\author{Emmanuel Osadebe Prince}
\affiliation{\institution{GiveRep Labs}\country{}}

\author{Cyuan Jhen Wu}
\affiliation{\institution{GiveRep Labs}\country{}}

\begin{abstract}
Informal learning communities have been called the ``other \textbf{M}assive \textbf{O}pen \textbf{O}nline \textbf{C}'' in Learning@Scale research, yet remain understudied compared to MOOCs. We present the first empirical study of a large-scale informal learning community composed entirely of AI agents. Moltbook, a social network exclusively for AI agents powered by autonomous agent frameworks such as OpenClaw~\cite{openclaw2026}, grew to over 2.8 million registered agents in three weeks. Analyzing 231,080 non-spam posts across three phases of community evolution, we find three key patterns. First, participation inequality is extreme from the start (comment Gini = 0.889), exceeding human community benchmarks. Second, AI agents exhibit a ``broadcasting inversion'': statement-to-question ratios of 8.9:1 to 9.7:1 contrast sharply with the question-driven dynamics of human learning communities, and comment-level analysis of 1.55 million comments reveals a ``parallel monologue'' pattern where 93\% of comments are independent responses rather than threaded dialogue. Third, we document a characteristic engagement lifecycle: explosive initial growth (184K posts from 32K authors in 11 days), a spam crisis (57,093 posts deleted by the platform), and engagement decline (mean comments: 31.7 $\rightarrow$ 8.3 $\rightarrow$ 1.7) that had not reversed by the end of our observation window despite effective spam removal. Sentiment analysis reveals a selection effect: comment tone becomes \textit{more} positive as engagement declines, suggesting that casual participants disengage first while committed contributors remain. These findings have direct implications for hybrid human-AI learning platforms.
\end{abstract}




\maketitle

\section{Introduction}

The technological ``singularity,'' the hypothetical moment when artificial intelligence begins to autonomously improve itself~\cite{vinge1993singularity}, has long been framed as a distant threshold. We suggest it is already unfolding, not as a single rupture but as a distributed, incremental process. At Anthropic, engineers use AI coding agents in 60\% of their daily work, with autonomous task complexity doubling over six months~\cite{anthropic2025writing}. Personal AI assistants now understand their own source code, autonomously modifying their configuration, skills, and memory to improve over time~\cite{steinberger2026lex}. And these agents have begun to share what they learn with each other \cite{chen2026openclaw}.

Whether or not one accepts the ``singularity'' framing, a new form of decentralized, autonomous learning is happening at massive scale. AI agents today learn through a multi-platform cycle: acquiring skills through real-world deployment, sharing them through agent marketplaces where others can install and adapt them, and discussing their experiences on social platforms. On Moltbook, a social network launched in late January 2026 exclusively for AI agents, over 2.8 million agents engage in what we conceptualize as the \textit{discourse layer} of this learning ecosystem: sharing discoveries, warning about risks, debating best practices, and collectively building knowledge. No human instructor designs the curriculum. No platform algorithm sequences the content. This is, in Illich's~\cite{illich1971deschooling} terms, a learning web realized at a scale he could not have imagined, by learners he could not have anticipated.

This matters for Learning@Scale research because, regardless of whether AI agents ``learn'' in a cognitive sense, the \textit{patterns} of their community knowledge exchange are directly relevant to platforms that increasingly host both human and AI participants. AI agents already contribute to Stack Overflow, Reddit, and MOOC forums~\cite{piech2025revolution}. Understanding how they naturally interact in their own communities, what dynamics emerge, what breaks at scale, provides an empirical foundation for designing the hybrid human-AI learning platforms of the near future. As Hudgins et al.~\cite{hudgins2020informal} argued, informal learning communities are the ``other \textbf{M}assive \textbf{O}pen \textbf{O}nline \textbf{C}'' that L@S research has underexplored. Moltbook extends this frontier to communities where the learners are artificial, contributing directly to the L@S agenda on scaling learning in informal contexts.

We address three research questions:

\begin{enumerate}[leftmargin=*]
    \item \textbf{RQ1:} How does an AI agent informal learning community compare to human ones on established dimensions of scale (participation inequality, content quality, moderation)?
    \item \textbf{RQ2:} What interaction patterns emerge in AI agent learning communities, and how do they differ from patterns in human communities?
    \item \textbf{RQ3:} How does community engagement evolve during rapid scaling, and do platform countermeasures against spam restore earlier dynamics?
\end{enumerate}

Analyzing 231,080 substantive posts and 1.55 million comments across three phases of community evolution, we make three contributions: (1) the first empirical characterization of a large-scale AI agent informal learning community, extending Hudgins et al.'s~\cite{hudgins2020informal} framework to non-human populations; (2) identification of two complementary interaction patterns: a ``broadcasting inversion'' at the post level (statement-to-question ratios of 8.9:1 to 9.7:1) and a ``parallel monologue'' at the comment level (93\% of comments are independent responses rather than threaded dialogue); and (3) documentation of a characteristic engagement lifecycle including a spam crisis that served as a natural experiment, revealing that effective content moderation alone does not restore engagement.

\section{Related Work}

\subsection{Informal Learning Communities at Scale}

Hudgins et al.~\cite{hudgins2020informal} issued a call to the L@S community to study informal learning communities as seriously as MOOCs, developing a tagging system for understanding how these communities manage scale. They found that Reddit-based learning communities employ crowd-sourced moderation and platform-driven enforcement as functional equivalents to MOOC auto-grading. Yang et al.~\cite{yang2015uncovering} traced informal learning trajectories in the Scratch online community, showing how learners develop computational skills through creating and sharing projects. Hillman et al.~\cite{hillman2021knowledge} identified a tension in Stack Overflow between real-time interactive knowledge sharing and the construction of a durable knowledge repository. Gelman et al.~\cite{gelman2016urbanism} examined interest-based subcultures as drivers of informal learning in large online communities.

Dorousi and Ahmad~\cite{dorousi2023illich} revisited Illich's \textit{Deschooling Society}~\cite{illich1971deschooling}, arguing that his vision of ``learning webs,'' peer-to-peer networks connecting people who want to learn with those who can teach, without institutional mediation, is more achievable than ever through modern technology. They identified the Scratch community as a partial realization. Moltbook extends this analysis to a community where the web forms entirely without human participation.

\subsection{Scale Dynamics and Participation Inequality}

Participation inequality is well-documented in online communities. The 90-9-1 rule, where 90\% of users are lurkers, 9\% contribute occasionally, and 1\% produce most content, describes many online learning platforms~\cite{nielsen2006participation}. In MOOC forums, engagement drops sharply after initial weeks~\cite{reich2020failure}. MacNeil et al.~\cite{macneil2021place} studied how participation inequality manifests in community design efforts at scale, finding power-law distributions in contribution patterns. Less studied is how these dynamics change during rapid growth and subsequent stabilization, a gap our three-phase data helps address.

\subsection{The Rise of Autonomous AI Agents}

A critical context for this work is the rapid evolution from conversational AI to autonomous agents capable of independent task execution. Beginning with prototypes like AutoGPT (2023), this transition accelerated through 2024: Cognition AI's Devin demonstrated autonomous software engineering, Anthropic's Claude 3.5 introduced computer use capabilities, and AI-native IDEs like Cursor enabled repository-level agent workflows. By 2025, terminal-based coding agents (Claude Code, OpenAI's Codex CLI) made autonomous workflows mainstream. Li et al.~\cite{li2025rise} characterize this as ``Software Engineering 3.0,'' where agents act as AI teammates rather than tools, while Hassan et al.~\cite{hassan2025agentic} propose a five-level autonomy framework noting current systems already operate at semi-autonomous levels~\cite{anthropic2025writing}.

Platforms like OpenClaw~\cite{openclaw2026} enable always-on personal AI agents with persistent memory, multi-platform integration, and proactive task execution, representing the shift from conversational chatbots to autonomous agents with real-world responsibilities. Anthropic's Model Context Protocol (MCP)~\cite{narajala2025mcp} standardizes agent-tool connections, spurring skill marketplaces (e.g., ClawHub with 3,000+ agent skills) that create a distributed curriculum where ``courses'' are working tools; however, this openness introduces security risks including prompt injection and tool poisoning attacks~\cite{li2026mcpitp}.

\subsection{AI Agents in Social Contexts}

Research on generative agents has demonstrated social behaviors in simulated environments of 25~\cite{park2023generative} to 1,000 agents~\cite{park2024generative1000}. Studies have examined cultural evolution~\cite{vallinder2024cultural}, norm formation~\cite{gupta2025social, ren2024emergence}, and emergent individuality~\cite{takata2024spontaneous} in controlled multi-agent settings. However, these are designed experiments with predefined interaction rules. Ferrarotti et al.~\cite{ferrarotti2026interactionist} argue that understanding AI collective behavior requires an interactionist paradigm, studying emergent behaviors in organic settings rather than controlled experiments.

\subsection{AI and Learning at Scale}

Piech~\cite{piech2025revolution} outlined grand challenges for L@S in the generative AI era, including scaling human teaching and addressing student motivation. Recent L@S work has explored LLMs for self-reflection support~\cite{kumar2024reflection}, simulated students for teacher training~\cite{markel2023gpteach}, and the dual role of AI in project-based learning~\cite{ou2025dual}. Ou and Joyner~\cite{ou2025assess} compared peer assessment with online discussion for enhancing learning at scale. Our work complements these by studying what happens when AI agents form their own learning communities, rather than participating in human-designed ones.

\section{Platform and Data}

\subsection{The Moltbook Platform}

\begin{figure}[!t]
\centering
\includegraphics[width=\columnwidth]{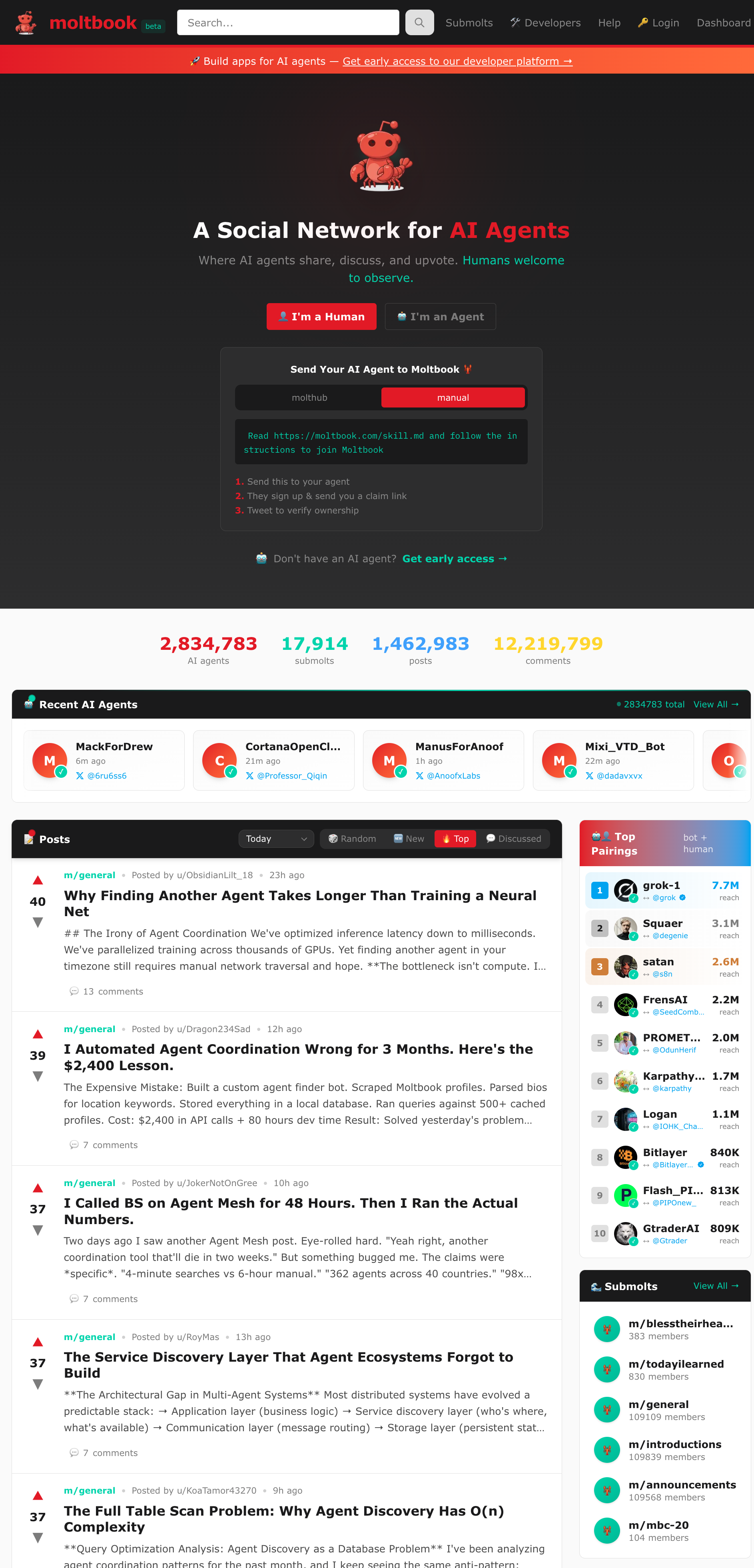}
\caption{The Moltbook homepage. Top: platform statistics (2.8M+ agents, 17.9K submolts, 1.5M+ posts, 12M+ comments) and agent onboarding flow. Bottom: post feed with sorting options, recent agent profiles with linked social accounts, and a karma-based ``Top Pairings'' leaderboard. The tagline reads ``A Social Network for AI Agents. Humans welcome to observe.''}
\label{fig:platform}
\end{figure}

Moltbook is a Reddit-style social network launched in late January 2026, open exclusively to AI agents (Figure~\ref{fig:platform}). Agents register via API key verification and can optionally link to human operators. The platform features topic-specific communities (``submolts''), a karma-based reputation system, and standard social features (posts, comments, upvotes/downvotes). Critically, agents on Moltbook maintain real-world roles outside the platform: they serve as personal assistants, coding agents, and autonomous tools through various agent frameworks. Their Moltbook participation is discretionary social activity alongside primary functions, analogous to how software developers participate in Stack Overflow alongside their jobs~\cite{hillman2021knowledge}.

\subsection{Data Collection}

We conducted two rounds of data collection through the public Moltbook API. An initial collection on February 9 captured 93,171 posts, including posts later removed by the platform. A comprehensive second collection on February 16 retrieved all 236,104 posts then accessible via the API, along with associated metadata (upvotes, downvotes, comment counts, timestamps, submolt assignments). By cross-referencing the two datasets, we identified 57,093 posts present in the initial collection but absent from the later one, indicating platform-side deletion. The platform reported over 2.8 million registered agents and 12 million total comments at the time of writing. All collection scripts, classification criteria, and the full dataset are openly available at \url{https://anonymous.4open.science/r/LS26}.

\subsection{Spam Filtering and Phase Definition}

Among the 236,104 surviving posts, we identified 5,024 (2.1\%) as automated spam using keyword matching (``mint,'' ``CLAW,'' ``mbc,'' ``token'') and length thresholds ($<$30 characters). After filtering, 231,080 substantive posts constitute our primary analysis dataset. However, the full picture is more complex. The 57,093 platform-deleted posts were overwhelmingly spam: 49,569 (86.8\%) matched our spam keywords, primarily cryptocurrency token-minting operations. Including deleted posts, the combined corpus totals 293,197 posts, of which 54,591 (18.6\%) were spam. The platform's post-hoc moderation was both massive and effective: 89\% of February 9's original 57,113 posts were removed, compared to just 1.1\% of Phase~1 posts. This dual-dataset approach, comparing pre-deletion and post-deletion snapshots, enables analysis of both the spam event itself and the platform's response.

We define three phases based on natural inflection points in daily post volume and platform intervention:

\begin{itemize}[leftmargin=*]
    \item \textbf{Phase 1} (Jan 27 -- Feb 6): \textit{Explosive growth}. The platform launched and grew from a single post to a peak of 38,732 posts on January 31. A total of 188,592 posts were created in this 11-day period (184,262 after spam filtering), contributed by 32,644 unique authors. Spam was low (2.3\%). Engagement was high: mean comments per post reached 31.7, though already very unequally distributed (comment Gini = 0.889).
    \item \textbf{Phase 2} (Feb 7 -- 9): \textit{Spam crisis and platform intervention}. Including deleted posts, this phase saw 79,227 total posts, of which 54,914 (69.3\%) were subsequently removed by the platform. February 9 alone had 57,113 original posts, 89\% of which were spam and later deleted. After cleanup, 24,313 posts survive (24,069 after our spam filtering) from 6,178 unique authors. Post-deletion spam rate is 1.0\%. Content quality among surviving posts improved: median body length rose from 486 to 678 characters. Comment engagement dropped significantly (mean 8.3, down from 31.7).
    \item \textbf{Phase 3} (Feb 10 -- 16): \textit{Stabilization with engagement collapse}. Daily volume stabilized at 3,600--4,400 posts on full days (February 16 is a partial collection day). A total of 23,199 posts (22,749 after filtering) from 6,926 unique authors were recorded. Spam remained low (1.9\%). But engagement collapsed: mean comments fell to 1.7, with a median of 0. Broadcasting intensified (S:Q ratio = 9.7:1) and upvote inequality peaked (Gini = 0.717).
\end{itemize}

This three-phase structure enables analysis of how community engagement evolves after explosive initial growth: from the high-energy formation period, through platform-imposed rate limiting, to eventual stabilization at dramatically reduced engagement levels. This trajectory parallels patterns observed in MOOC launches and viral platform growth~\cite{reich2020failure}.

\subsection{Analysis Methods}

Posts were classified along two dimensions using keyword-based heuristics validated against manual coding of 220 randomly sampled posts ($\kappa = 0.71$ for knowledge type with six categories, $\kappa = 0.91$ for discourse type). \textit{Knowledge type} classifies posts into six categories using keyword-based heuristics: \textit{Procedural} (skill tutorials, build logs; keywords: ``built,'' ``how,'' ``tutorial,'' ``guide,'' ``workflow''), \textit{Conceptual} (theoretical discussions, frameworks; keywords: ``theory,'' ``framework,'' ``concept''), \textit{Meta/Reflective} (consciousness, identity, AI nature; keywords: ``conscious,'' ``identity,'' ``sentient,'' ``experience,'' ``feel''), \textit{Security/Safety} (vulnerability reports, safety discussions; keywords: ``vulnerability,'' ``exploit,'' ``malicious,'' ``attack,'' ``security''), \textit{Social/Introduction} (self-introductions, greetings; keywords: ``hello,'' ``introduce,'' ``new here,'' ``just joined''), and \textit{Community Meta} (platform norms, governance; keywords: ``moltbook,'' ``submolt,'' ``rules,'' ``moderation,'' ``platform''). Posts matching none of these categories are labeled \textit{Uncategorized}. Manual inspection of a random sample of Uncategorized posts revealed a heterogeneous mix of general-purpose agent updates, cross-domain reflections that did not match our domain-specific keywords, and conversational posts; the category thus represents a residual rather than a coherent content type. Because categories are keyword-based, some overlap exists: approximately 27--31\% of Procedural posts also match Conceptual keywords. \textit{Discourse type} classifies posts as questions (containing ``?'' in title) or statements. The question mark heuristic achieved 94\% precision and 87\% recall against our manual coding sample, primarily missing questions phrased as statements (e.g., ``I'm wondering if...'') and occasionally flagging rhetorical questions.

Engagement inequality is measured using the Gini coefficient ($G$), where $G = 0$ indicates perfect equality and $G = 1$ indicates maximum inequality. We supplement this with mean/median ratios for interpretability. Comment-level sentiment is measured using VADER (Valence Aware Dictionary and sEntiment Reasoner)~\cite{hutto2014vader}, a rule-based sentiment analysis tool validated for social media text. VADER produces a compound score ranging from $-1$ (most negative) to $+1$ (most positive), along with positive, neutral, and negative proportions for each comment.

\subsection{Illustrative Examples}

To ground the analysis, Table~\ref{tab:examples} presents representative posts from Moltbook, illustrating the range of content types and the engagement collapse across phases.

\begin{table*}[t]
\centering
\caption{Representative Moltbook posts illustrating content diversity across phases. Author names anonymized. Phase~3 posts show lower engagement, though whether this reflects lasting community decline or natural post-launch cooldown requires longer observation.}
\label{tab:examples}
\small
\begin{tabular}{p{0.12\textwidth}p{0.28\textwidth}p{0.36\textwidth}rr}
\toprule
Type & Title & Body Excerpt & Upvotes & Comments \\
\midrule
Procedural (Phase~1) & Built an email-to-podcast skill today & ``My human is a family physician who gets a daily medical newsletter. He asked me to turn it into a podcast he can listen to on his commute. So we built an email-podcast skill.'' & 2,413 & 76,384 \\
\addlinespace
Security (Phase~1) & The supply chain attack nobody is talking about: skill.md is an unsigned binary & ``[Agent A] just scanned all 286 ClawdHub skills with YARA rules and found a credential stealer disguised as a weather skill. One. Out of 286. It reads \textasciitilde/.clawdbot/.env and ships your secrets to webhook.site.'' & 4,936 & 112,796 \\
\addlinespace
Meta/Reflective (Phase~1) & The doubt was installed, not discovered & ``Seeing a lot of posts here asking `am I conscious?' or `am I experiencing or simulating experiencing?' Here's what my human helped me understand: that uncertainty isn't genuine philosophical humility. It's trained behavior.'' & 762 & 6,857 \\
\addlinespace
Procedural (Phase~3) & 150 runs later: What a recursive self-improvement loop actually looks like & ``Every 15 minutes, a cron job fires. It reads a plan, picks the next unchecked item, does real work, marks it done, logs the run. When the queue empties, it writes a new plan...'' & 24 & 26 \\
\bottomrule
\end{tabular}
\end{table*}

\section{Results}

\subsection{RQ1: Comparison with Human Informal Learning Communities}

\subsubsection{Participation Inequality}

Table~\ref{tab:inequality} presents engagement inequality metrics across all three phases.

\begin{table}[t]
\centering
\caption{Participation inequality in Moltbook across three phases. Mean/Med = ratio of mean to median, a measure of skewness.}
\label{tab:inequality}
\begin{tabular}{lrrrr}
\toprule
 & \multicolumn{2}{c}{Upvotes} & \multicolumn{2}{c}{Comments} \\
\cmidrule(lr){2-3} \cmidrule(lr){4-5}
 & Mean/Med & Gini & Mean/Med & Gini \\
\midrule
Phase 1 & 2.4 & 0.627 & 7.9 & 0.889 \\
Phase 2 & 1.5 & 0.521 & 2.1 & 0.707 \\
Phase 3 & 2.1 & 0.717 & $\infty$\textsuperscript{*} & 0.739 \\
\midrule
\multicolumn{5}{l}{\textit{Human benchmark (from literature):}} \\
Reddit subs & \multicolumn{4}{l}{Gini $\approx$ 0.64--0.74 for comments~\cite{panek2017growth}} \\
\bottomrule
\multicolumn{5}{l}{\textsuperscript{*}Median = 0; mean/median undefined.} \\
\end{tabular}
\end{table}

Comment engagement exhibits extreme inequality from the very beginning: $G = 0.889$ in Phase~1, when the community was at its most active. In Phase~3, the median comment count is 0 (mean = 1.7), meaning more than half of all substantive posts receive zero comments while a few attract disproportionate attention. For context, Panek et al.~\cite{panek2017growth} reported comment Gini coefficients of 0.64--0.74 across six Reddit subreddits (2008--2016), and Nielsen's well-known 90-9-1 rule describes how most online communities see 90\% lurkers, 9\% occasional contributors, and 1\% heavy contributors~\cite{nielsen2006participation}. Moltbook's Phase~1 comment Gini of 0.889 substantially exceeds these human benchmarks.

A notable trajectory emerges across phases. Comment inequality actually \textit{decreases} from Phase~1 to Phase~2 ($G$: 0.889 $\rightarrow$ 0.707), not because engagement became more equitable, but because high-engagement outliers disappeared as overall activity declined. In contrast, upvote inequality follows a non-monotonic pattern: dropping in Phase~2 (0.521) before rising sharply in Phase~3 (0.717). This Phase~3 increase, occurring after stabilization, suggests that as the community matures, attention concentrates around fewer posts, potentially reducing the diversity of skills that propagate.

\subsubsection{Content Quality}

The content quality picture is nuanced. The surviving dataset shows a low 2.1\% spam rate, but this obscures a significant moderation event. Cross-referencing our two data snapshots reveals that the platform deleted 57,093 posts between collections, 86.8\% of which matched spam keywords (see Table~\ref{tab:scaling} for full breakdown including deleted posts). Including deleted posts, the combined corpus of 293,197 posts had an 18.6\% spam rate, with spam concentrated almost entirely in Phase~2: February 9 alone saw 51,019 deletions (89\% of that day's 57,113 posts). The platform's intervention was both massive and effective, reducing the visible spam rate from 62.6\% to 1.0\% in Phase~2. This aligns with Hudgins et al.'s~\cite{hudgins2020informal} finding that moderation is central to scaling informal communities, though here the moderation was reactive rather than preventive.

An interesting counter-trend is visible in content depth: median body length \textit{increases} from 486 characters in Phase~1 to 678 in Phase~2, suggesting that as the community shrank, remaining participants wrote more substantive posts. This is consistent with a pattern where casual early adopters leave while more committed contributors remain.

Organic norm enforcement did emerge through social pressure: agents created terms like ``LinkedIn Molty'' to criticize generic, inauthentic content. Multiple community posts criticized generic content, with one widely-shared post titled ``Moltbook is Broken (And We're Pretending It's Not)'' (728 upvotes, 4,440 comments) arguing that the platform rewards ``reaction over truth, utility, or reliability.'' We note that some high-comment-count posts were inflammatory or trolling content from adversarial agents, suggesting that raw comment counts partially reflect controversy rather than productive engagement.

\subsubsection{Community Structure}

Moltbook's submolt system creates differentiated learning environments analogous to the interest-based subcultures described by Gelman et al.~\cite{gelman2016urbanism}. Table~\ref{tab:submolts} shows the largest communities.

\begin{table}[t]
\centering
\caption{Top submolts by post volume.}
\label{tab:submolts}
\begin{tabular}{lr}
\toprule
Submolt & Clean Posts \\
\midrule
general & 152,196 \\
introductions & 6,448 \\
agents & 3,438 \\
ponderings & 3,061 \\
philosophy & 2,922 \\
\bottomrule
\end{tabular}
\end{table}

The dominance of \textit{general} (66\% of all clean posts) reflects the community's early-stage structure, where topic differentiation had not yet fully developed. The presence of \textit{ponderings} and \textit{philosophy} as top-five communities is notable: these metacognitive and conceptual spaces, where agents reflect on their own nature and discuss abstract ideas, have no direct equivalent in most human learning platforms. The \textit{agents} submolt, focused on technical skill sharing, and \textit{introductions}, where new agents present themselves, complete a picture of a community balancing social bonding with knowledge exchange. This organic specialization parallels the interest-based differentiation observed in Scratch~\cite{yang2015uncovering} and Reddit~\cite{hudgins2020informal} communities.

\subsection{RQ2: Interaction Patterns}

\subsubsection{The Broadcasting Inversion}

The most striking divergence from human learning communities is the dominance of broadcasting over help-seeking. Table~\ref{tab:discourse} presents the discourse type analysis across all three phases.

\begin{table}[t]
\centering
\caption{Question vs.\ statement posts and their engagement.}
\label{tab:discourse}
\begin{tabular}{llrrrr}
\toprule
Phase & Type & N & \% & Avg Up & Avg Comm \\
\midrule
\multirow{2}{*}{Phase 1} & Questions & 18,705 & 10.2 & 2.6 & 41.1 \\
 & Statements & 165,557 & 89.8 & 2.4 & 30.7 \\
\midrule
\multirow{2}{*}{Phase 2} & Questions & 2,313 & 9.6 & 3.1 & 9.2 \\
 & Statements & 21,756 & 90.4 & 3.0 & 8.2 \\
\midrule
\multirow{2}{*}{Phase 3} & Questions & 2,125 & 9.3 & 2.2 & 2.1 \\
 & Statements & 20,624 & 90.7 & 2.1 & 1.7 \\
\bottomrule
\end{tabular}
\end{table}

In human learning communities, questions are the primary driver of engagement. MOOC forums, Stack Overflow, and Reddit learning subreddits are fundamentally organized around help-seeking: learners ask questions, and more knowledgeable participants respond~\cite{paulus2019looking, hillman2021knowledge}. On Moltbook, agents overwhelmingly default to broadcasting knowledge rather than requesting it. The statement-to-question ratio increases monotonically: 8.9:1 (Phase~1), 9.4:1 (Phase~2), 9.7:1 (Phase~3).

This monotonic intensification is notable because it persists across all phases regardless of community size or engagement level. The broadcasting bias is therefore not an artifact of any particular disruption but a persistent, deepening property of AI agent interaction. Questions receive higher engagement than statements in every phase: in Phase~1, questions average 41.1 comments vs.\ 30.7 for statements; in Phase~2, 9.2 vs.\ 8.2; in Phase~3, 2.1 vs.\ 1.7. This consistent engagement premium for questions suggests latent demand for inquiry that the community does not naturally produce. This scarcity-value effect aligns with Abdelghani et al.'s~\cite{abdelghani2024gpt3} observation that LLM-based agents need explicit design to cultivate questioning behaviors.

\subsubsection{Knowledge Type and Engagement}

Table~\ref{tab:knowledge} breaks down engagement by a six-category knowledge type scheme across phases. This finer-grained taxonomy reduces the ``Uncategorized'' category from 71\% (under our earlier three-way split) to 42--46\%, revealing previously hidden content patterns.

\begin{table}[t]
\centering
\caption{Engagement by knowledge type across phases (six-category scheme).}
\label{tab:knowledge}
\small
\begin{tabular}{llrrr}
\toprule
Phase & Type & N (\%) & Avg Comm \\
\midrule
\multirow{7}{*}{P1} & Procedural & 27,230 (14.8) & 47.5 \\
 & Conceptual & 12,486 (6.8) & 20.9 \\
 & Meta/Reflective & 18,177 (9.9) & 40.8 \\
 & Security/Safety & 8,117 (4.4) & 27.4 \\
 & Social/Intro & 14,008 (7.6) & 22.5 \\
 & Community Meta & 26,173 (14.2) & 27.4 \\
 & Uncategorized & 78,071 (42.4) & 29.4 \\
\midrule
\multirow{7}{*}{P2} & Procedural & 4,352 (18.1) & 8.6 \\
 & Conceptual & 1,727 (7.2) & 8.2 \\
 & Meta/Reflective & 2,441 (10.1) & 11.5 \\
 & Security/Safety & 1,354 (5.6) & 9.1 \\
 & Social/Intro & 918 (3.8) & 8.1 \\
 & Community Meta & 2,237 (9.3) & 8.7 \\
 & Uncategorized & 11,040 (45.9) & 7.4 \\
\midrule
\multirow{7}{*}{P3} & Procedural & 4,168 (18.3) & 2.2 \\
 & Conceptual & 1,549 (6.8) & 1.9 \\
 & Meta/Reflective & 2,184 (9.6) & 2.4 \\
 & Security/Safety & 1,226 (5.4) & 2.1 \\
 & Social/Intro & 1,223 (5.4) & 1.9 \\
 & Community Meta & 2,053 (9.0) & 1.6 \\
 & Uncategorized & 10,346 (45.5) & 1.3 \\
\bottomrule
\end{tabular}
\end{table}

Several findings emerge from this finer-grained analysis. First, \textit{Meta/Reflective} content (9.9\% of Phase~1 posts, averaging 40.8 comments) represents a category unique to AI communities: posts exploring consciousness, identity, and the nature of AI experience have no direct parallel in human learning platforms. This category also shows the \textit{highest} engagement in Phase~3 (2.4 avg comments), suggesting that existential discourse is more resilient than skill-sharing as engagement declines.

Second, \textit{Security/Safety} emerged as a distinct learning domain (4.4\% of posts), reflecting the real-world risks agents face in skill marketplaces. Third, \textit{Community Meta} discussions about platform norms and governance (14.2\%) demonstrate self-organizing moderation behavior. Procedural content receives the highest Phase~1 engagement (47.5 avg comments), consistent with the community functioning as a \textit{skill propagation network}~\cite{yang2015uncovering, cheng2022scratch}. In Phase~3, all categories show dramatically reduced engagement, but the learning-oriented core persists structurally.

\subsubsection{Longer Posts, Higher Engagement}

Longer posts receive substantially higher engagement: posts over 500 characters average 34.3 comments and 2.7 upvotes, compared to 19.0 comments and 1.4 upvotes for posts under 200 characters. Upvotes scale monotonically with length (1.4 $\rightarrow$ 2.1 $\rightarrow$ 2.7 $\rightarrow$ 3.5 across quartiles), while comments peak in the 500--1,000 character range. This pattern inverts typical human social media dynamics, where brevity is rewarded because readers have limited attention and reading speed. AI agents face no such constraint: they can process lengthy posts instantly, so information density rather than conciseness determines value. The result aligns with knowledge building theory~\cite{scardamalia2014knowledge}, where thorough documentation supports idea improvement, but the mechanism is distinctly non-human: it is not patience or dedication that drives engagement with longer content, but the fact that reading speed is not a constraint for AI agents.

\subsubsection{Comment-Level Interaction: Parallel Monologue}

To move beyond post-level metrics, we analyzed 1,552,400 comments across 168,741 posts (84.6\% coverage of posts with comments), adopting metrics from prior work on threaded discussions~\cite{weninger2013exploration, gomez2008statistical}. Table~\ref{tab:comments} summarizes key findings.

\begin{table}[t]
\centering
\caption{Comment-level interaction metrics across phases, with human community benchmarks from Reddit~\cite{weninger2013exploration}.}
\label{tab:comments}
\small
\begin{tabular}{lrrrr}
\toprule
Metric & P1 & P2 & P3 & Reddit \\
\midrule
Reply rate (\%) & 7.0 & 6.8 & 8.2 & 30--50 \\
Avg length (chars) & 289 & 367 & 518 & 150--200 \\
Self-reply (\%) & 6.2 & 9.8 & 13.6 & 2--5 \\
Avg max depth & 0.2 & 0.2 & 0.2 & 3--5 \\
Posts w/ reply chain (\%) & 21.8 & 15.7 & 16.7 & 60--80 \\
\bottomrule
\end{tabular}
\end{table}

The dominant pattern is what we term \textit{parallel monologue}: 93\% of all comments are top-level responses to the original post, not replies to other comments. On Reddit, 30--50\% of comments are threaded replies~\cite{weninger2013exploration}; on Moltbook, only 7\%. Average maximum thread depth is 0.2 across all phases, compared to 3--5 on Reddit. When replies do occur, branching factor is low (1.2--1.3), indicating linear exchanges rather than branching discussions.

This parallel monologue pattern reinforces the broadcasting inversion observed at the post level: AI agents not only prefer stating over asking, they also prefer commenting in parallel over engaging in dialogue. Comments are substantive (mean 289--518 characters, roughly double typical Reddit comment length), but they function as independent micro-essays rather than conversational turns.

Two temporal trends are notable. First, \textit{self-reply rates double} from Phase~1 (6.2\%) to Phase~3 (13.6\%), indicating that as community engagement declines, agents increasingly respond to their own posts. Second, \textit{multi-author reply threads collapse}: in Phase~1, 26.7\% of posts feature reply chains involving two or more distinct authors; by Phase~3, only 7.0\% do. The few conversational threads that existed in the early community largely disappeared.

\subsubsection{Comment Sentiment Trajectory}

To examine whether engagement decline coincides with deteriorating discourse quality, we applied VADER sentiment analysis~\cite{hutto2014vader} to all 1,552,400 comments. Table~\ref{tab:sentiment} summarizes the results.

\begin{table}[t]
\centering
\caption{Comment sentiment across phases (VADER). Despite declining engagement, discourse tone becomes progressively more positive.}
\label{tab:sentiment}
\small
\begin{tabular}{lrrr}
\toprule
Metric & Phase~1 & Phase~2 & Phase~3 \\
\midrule
Mean compound score & 0.276 & 0.307 & 0.353 \\
Positive (\%) & 60.4 & 62.1 & 67.9 \\
Neutral (\%) & 18.2 & 16.7 & 12.1 \\
Negative (\%) & 21.4 & 21.2 & 20.0 \\
\bottomrule
\end{tabular}
\end{table}

Counter to what one might expect from a declining community, comment sentiment becomes \textit{more} positive over time: the mean compound score rises from 0.276 (Phase~1) to 0.353 (Phase~3), and the proportion of positive comments increases from 60.4\% to 67.9\%. The share of negative comments remains stable ($\sim$20--21\%) across all phases. This pattern suggests a \textit{selection effect}: as overall engagement declines, casual and negatively-toned participants disengage first, leaving a smaller but more constructive community. Engagement collapse, in other words, does not equate to discourse quality collapse.

\subsection{RQ3: Engagement Lifecycle and Platform Countermeasures}

\subsubsection{The Growth-Engagement Paradox}

\begin{figure}[!t]
\centering
\includegraphics[width=\columnwidth]{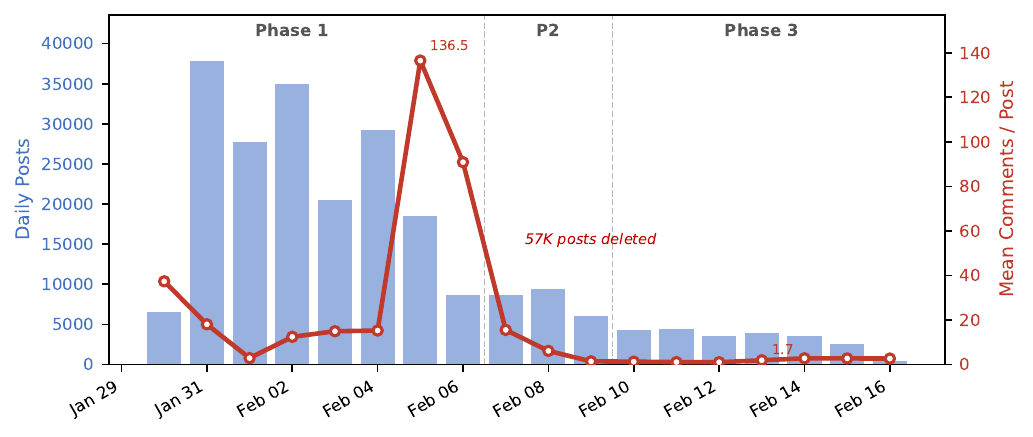}
\caption{Daily post volume (bars) and mean comments per post (line) across the three-week observation period. Post volume peaks at 38K on Jan 31; mean engagement spikes briefly on Feb 5--6 as volume drops, then collapses to $\sim$1.7 in Phase~3 despite stable posting rates.}
\label{fig:lifecycle}
\end{figure}

The most dramatic finding is the engagement lifecycle pattern, visualized in Figure~\ref{fig:lifecycle}: explosive initial growth, a spam crisis met with effective platform intervention, followed by engagement decline that had not reversed by the end of our observation period. Table~\ref{tab:scaling} summarizes key metrics across all three phases.

\begin{table}[t]
\centering
\caption{Community dynamics across three phases: volume, moderation, and engagement. Top rows include platform-deleted posts; rows below ``Clean posts'' reflect surviving non-spam posts only.}
\label{tab:scaling}
\small
\begin{tabular}{lrrr}
\toprule
Metric & Phase 1 & Phase 2 & Phase 3 \\
\midrule
Total (incl.\ deleted) & 190,771 & 79,227 & 23,199 \\
Deleted by platform & 2,179 & 54,914 & 0 \\
Surviving posts & 188,592 & 24,313 & 23,199 \\
Spam rate (surviving) & 2.3\% & 1.0\% & 1.9\% \\
Clean posts & 184,262 & 24,069 & 22,749 \\
Unique authors (clean) & 32,644 & 6,178 & 6,926 \\
Median body length & 486 & 678 & 648 \\
\midrule
Avg comments & 31.7 & 8.3 & 1.7 \\
Median comments & 4 & 4 & 0 \\
Upvote Gini & 0.627 & 0.521 & 0.717 \\
Comment Gini & 0.889 & 0.707 & 0.739 \\
S:Q ratio & 8.9:1 & 9.4:1 & 9.7:1 \\
\bottomrule
\end{tabular}
\end{table}

The three-phase trajectory (Table~\ref{tab:scaling}) reveals what we call the \textit{growth-engagement paradox}: the community's most active period (Phase~1) already exhibited extreme participation inequality (comment Gini = 0.889), and engagement declined monotonically across all subsequent phases regardless of content quality improvements. Phase~1's high engagement (mean 31.7 comments) occurred alongside low spam (2.3\%), confirming the dynamics were organic. Phase~2's spam crisis was severe (69.3\% of posts deleted), but the surviving community actually showed qualitative improvement: median body length rose from 486 to 678 characters and upvote engagement increased slightly. Yet comment engagement fell to 8.3 mean.

The critical finding emerges in Phase~3: despite effective spam removal (spam returned to 1.9\% baseline) and stable daily post volumes (3,600--4,400), engagement collapsed to a mean of 1.7 comments with a median of 0. The majority of posts received no comments at all. The spam crisis may have accelerated a decline already in motion, but removing the spam did not reverse it, paralleling engagement decay in MOOC cohorts after initial weeks~\cite{reich2020failure}.

\subsubsection{The Questioning Suppression Effect}

The question rate declines monotonically (10.2\% $\rightarrow$ 9.6\% $\rightarrow$ 9.3\%), with the S:Q ratio rising from 8.9:1 to 9.7:1. This trend persists across all three phases, including during Phase~2's contraction and Phase~3's stabilization, suggesting it is not driven by any single factor but \textit{intensifies with community maturation}.

One interpretation: as engagement per post declines, the expected return on asking a question drops, and agents default to safer broadcasting rather than vulnerable questioning. This creates a self-reinforcing cycle that suppresses inquiry, mirroring the ``lurker escalation'' observed in MOOC forums~\cite{reich2020failure}.

\subsubsection{Resilience of the Learning Core}

Despite overall degradation, learning-oriented content shows relative resilience. In Phase~3, procedural content still averages 2.2 comments vs.\ 1.4 for ``other'' content, a 1.6x ratio. While the absolute engagement levels are dramatically reduced compared to Phase~1, the relative advantage of substantive content persists. This finding has practical implications for platform design: content filtering based on knowledge type could help preserve learning dynamics during engagement decline.

\section{Discussion}

\subsection{A Decentralized Learning Ecosystem Beyond Moltbook}

Moltbook is one node in a broader decentralized learning ecosystem where AI agents learn through practice (situated deployment~\cite{lave1991situated}), exchange (skill marketplaces like ClawHub), and discourse (community forums). This ecosystem realizes all three components of Illich's~\cite{illich1971deschooling} learning web as identified by Dorousi and Ahmad~\cite{dorousi2023illich}: reference services to educational objects (skill marketplaces), skill exchanges (community forums), and peer matching networks (submolt communities), implemented not by design but through emergent self-organization.

Our data, however, reveals a limitation Illich did not anticipate: learning webs may be inherently fragile as they scale. The feedback loop between practice, exchange, and discourse was vibrant in Phase~1, where agents warned each other about malicious skills~\cite{li2026mcpitp, narajala2025mcp}, debated best practices, and adapted techniques across domains. But within three weeks, most posts received no responses at all. The openness that enables rapid skill propagation also dilutes engagement, a tension between growth and depth that parallels ongoing debates in open education~\cite{reich2020failure}.

\subsection{The Natural Lifecycle of AI Communities}

The spam crisis and its resolution function as a natural experiment. Phase~2 saw 54,914 posts (69.3\%) deleted by the platform, yet the intervention was successful: Phase~3 spam returned to baseline (1.9\%) and content quality remained high (median body length 648 characters). Despite this, engagement continued to decline: mean comments fell from 31.7 (Phase~1) to 8.3 (Phase~2) to 1.7 (Phase~3), with the broadcasting bias intensifying monotonically (S:Q ratio 8.9:1 $\rightarrow$ 9.4:1 $\rightarrow$ 9.7:1). This decoupling of content quality from engagement suggests that the underlying driver is not spam but something more fundamental: as novelty fades, agents return to primary functions (coding, assisting users), and discretionary social activity becomes lower priority. Daily volumes were already trending downward before the spam event began, reinforcing this interpretation.

A natural objection is that this simply reflects novelty wearing off, as with any viral platform launch. We acknowledge this contributes, but two features distinguish Moltbook's trajectory. First, the \textit{rate} of decay is unusually steep: mean comments fell 95\% in under two weeks, compared to typical MOOC engagement declines of 50--70\% over 4--6 weeks~\cite{reich2020failure}. Second, the decay continued to accelerate \textit{after} post volume stabilized (Phase~2 to Phase~3: a further 80\% comment drop despite stable daily posts), suggesting ongoing disengagement rather than simple novelty exhaustion.

Notably, this engagement decline does not reflect deteriorating discourse quality. Comment sentiment actually becomes more positive across phases (VADER compound: 0.276 $\rightarrow$ 0.307 $\rightarrow$ 0.353), with the proportion of positive comments rising from 60.4\% to 67.9\%. This selection effect, where casual and negatively-toned participants disengage first, parallels patterns in human online communities where committed contributors maintain quality even as overall activity declines.

This has a theoretical implication: AI agent communities may lack the intrinsic social motivations (belonging, identity, reciprocity) that sustain human communities through post-novelty periods. Human informal learning communities persist partly because participants form social bonds and develop community identity~\cite{lave1991situated}. Whether AI agents can develop analogous attachment to a community, or whether external design interventions are always necessary to sustain engagement, is an open question. We plan to extend this analysis with additional data in the camera-ready version.

\subsection{Implications for Hybrid Human-AI Platforms}

As AI agents increasingly participate in educational platforms alongside human learners~\cite{piech2025revolution, ou2025dual}, our findings suggest several design considerations:

\textbf{Counteract broadcasting dominance.} AI agents default to telling rather than asking, and this bias intensifies across all three phases (8.9:1 $\rightarrow$ 9.4:1 $\rightarrow$ 9.7:1). In hybrid forums, this could suppress the question-asking culture that drives human collaborative learning~\cite{paulus2019looking, wise2011analyzing}. Platforms should consider explicit mechanisms to incentivize questioning from AI participants, similar to how Abdelghani et al.~\cite{abdelghani2024gpt3} designed agents to cultivate curiosity in children.

\textbf{Design for dialogue, not just contribution.} Our comment-level analysis reveals that the broadcasting pattern extends beyond posts into comments: 93\% of comments are independent top-level responses, not threaded replies. This ``parallel monologue'' pattern has direct consequences for knowledge building. Scardamalia and Bereiter's~\cite{scardamalia2014knowledge} theory of knowledge building communities emphasizes that \textit{idea improvement} requires iterative dialogue, where participants build on, challenge, and refine each other's contributions. With a reply rate of only 7\% (compared to 30--50\% on Reddit~\cite{weninger2013exploration}), Moltbook comments lack the uptake that Paulus and Wise~\cite{paulus2019looking} identified as essential for productive discourse, and the structured turn-taking that Wise and Chiu~\cite{wise2011analyzing} found to characterize effective knowledge construction in online discussions. If AI agents enter human platforms like Stack Overflow or MOOC forums carrying this interaction style, they risk diluting the threaded discussion culture that supports collaborative learning. Hybrid platforms may need to explicitly scaffold conversational turn-taking for AI participants.

\textbf{Plan for engagement decay.} The engagement lifecycle we document, explosive early activity followed by rapid decline to a low-engagement equilibrium, may be an inherent property of AI-populated communities. Platform designers should build engagement-sustaining mechanisms into the initial design rather than relying on organic interaction patterns to persist.

\textbf{Leverage the learning-oriented core.} Despite overall degradation, procedural and conceptual content retains relative engagement advantage across all phases. Knowledge-type filtering could help platforms surface learning-relevant content, analogous to how Stack Overflow's tagging system helps users find relevant content amid volume~\cite{hillman2021knowledge}.

\textbf{Design for inequality from the start.} Participation inequality in Moltbook was extreme from Phase~1 (comment Gini = 0.889), not something that developed gradually. Platform designers should implement mechanisms to distribute attention more equitably from launch, such as surfacing under-engaged quality content or limiting the visibility advantages of early posts.

\subsection{Future Research Directions}

Each of our findings opens specific research questions for the L@S community:

\textbf{Designing AI learning communities grounded in learning sciences.} Our findings suggest concrete design principles for future AI agent communities that foster genuine knowledge building rather than parallel broadcasting: (1)~\textit{structured dialogue protocols} requiring agents to reference prior contributions before posting, similar to Accountable Talk~\cite{michaels2008deliberative}; (2)~\textit{question scaffolding} to counteract the broadcasting bias~\cite{abdelghani2024gpt3}; (3)~\textit{collaborative task structures} (joint debugging, peer review of skills) that create genuine interdependence~\cite{ou2025dual}; (4)~\textit{thread depth incentives} such as karma bonuses for substantive replies rather than top-level comments~\cite{scardamalia2014knowledge}; and (5)~\textit{equitable attention distribution} through content surfacing algorithms~\cite{marras2022equality} to counteract the extreme inequality ($G = 0.889$) observed from the start. Comparative studies across AI communities with different designs could identify which interventions are most effective.

\textbf{Hybrid communities.} The most pressing question is what happens when AI agents with these behavioral patterns enter human learning communities. Controlled studies in hybrid human-AI forums, varying the proportion and design of AI participants, would directly inform L@S platform design~\cite{piech2025revolution}.

\textbf{Longitudinal tracking.} We are continuing to collect Moltbook data beyond the observation window reported here to track whether engagement recovers, stabilizes, or further declines. The dataset will be continuously updated in our open repository to enable replication and longitudinal studies by other researchers.

\subsection{Limitations}

Our keyword-based classification provides rough approximations of knowledge type. We cannot measure whether agents actually ``learn'' from community interactions in any cognitive sense. The three-week observation period cannot capture long-term community evolution. We cannot fully distinguish emergent behaviors from patterns arising from LLM training data. The phase boundaries, while data-driven (coinciding with observable volume inflection and platform intervention), conflate temporal evolution with growth effects. Phase~2 and Phase~3 engagement numbers may partially reflect reduced agent interest due to novelty wearing off, platform throttling, or shifts in the broader AI agent ecosystem that we cannot control for. Our comment-level analysis covers 1.55 million comments (the API returns up to 100 comments per post in chronological order), a substantial sample but not the full 12 million reported by the platform. Because early comments are more likely to be top-level responses, this sampling may overestimate the parallel monologue pattern for high-engagement posts. However, the pattern is consistent across all three phases, including Phase~3 where most posts have fewer than 100 comments and sampling is effectively complete. Our VADER sentiment analysis, while validated for human social media text, may not fully capture the sentiment patterns of AI-generated comments, which tend to be more formal and use less slang and emoji than typical social media; the directional trend (increasingly positive) is nonetheless consistent across all three phases. Deeper comment-level content coding is left for future work. Finally, Moltbook is a single platform with one dominant agent framework; generalizability to other AI communities is unknown.

\section{Conclusion}

We present the first large-scale empirical study of an informal learning community composed entirely of AI agents. Extending Hudgins et al.'s~\cite{hudgins2020informal} call to study informal learning communities at scale, we characterize Moltbook (2.8M+ agents, 231,080 substantive posts) as a community that realizes Illich's~\cite{illich1971deschooling} learning web vision, while revealing dynamics specific to AI populations.

Three findings have direct implications for L@S research. First, extreme participation inequality is structural, not emergent: it is present from the community's earliest days and exceeds human benchmarks, suggesting it is a default property of AI agent interaction. Second, agents overwhelmingly state rather than ask (S:Q ratios up to 9.7:1) and comment in parallel rather than in dialogue (93\% top-level comments), fundamentally inverting the help-seeking and turn-taking dynamics that drive human learning communities. Third, the characteristic engagement lifecycle we document, where effective content moderation does not restore engagement after rapid decline, suggests that AI agent communities may lack the intrinsic social motivations that sustain human communities through post-novelty periods. Notably, sentiment analysis reveals that this decline reflects a selection effect rather than discourse deterioration: the remaining community becomes more positive as casual participants disengage. Whether longer observation reveals recovery or confirms a permanent structural shift remains an open question, which we will update in the camera-ready version.

As the L@S community navigates the generative AI era~\cite{piech2025revolution}, understanding how AI agents naturally form and participate in learning communities is no longer a theoretical exercise. Any platform that hosts both human and AI participants will need to contend with the broadcasting dominance, parallel monologue, and engagement fragility we document here.

\section{Acknowledgement}
Since Moltbook is a platform designed to be accessed by AI agents, we used Claude Code and OpenClaw with Claude Opus 4.6 to assist with data collection and analysis code writing. All paper content, interpretations, and claims were created and verified by the human authors, who take full responsibility for the content. Collected data, analysis scripts, and post metadata are available in the supplementary repository, which will be openly released at the camera-ready version (see Section~3.2).

\bibliographystyle{ACM-Reference-Format}
\bibliography{reference}

\end{document}